\documentclass{PoS}

\usepackage{graphics}
\usepackage{bm}
\usepackage{bbm}
\usepackage{yfonts}
\usepackage{array}
\usepackage{amssymb}
\usepackage{amsopn}
\usepackage{amsmath} 
\usepackage{mathrsfs}
\usepackage{lmodern}
\usepackage{url}

\title{Computing the full two-loop gluon Regge trajectory within Lipatov's high energy effective action}

\ShortTitle{NLO gluon Regge trajectory and Lipatov's effective action}

\author{\speaker{G. Chachamis}\\
    Instituto de F\'{\i}sica Corpuscular, Universitat de Val\`encia -- 
Consejo Superior de Investigaciones Cient\'{\i}ficas,
Parc Cient\'{\i}fic, E-46980 Paterna (Valencia), Spain\\  
        E-mail: \email{grigorios.chachamis@ific.uv.es}}

\author{M.~Hentschinski\\
        Physics Department, Brookhaven National Laboratory, Upton, NY 11973, USA\\
        E-mail: \email{hentsch@bnl.gov}}

\author{J.~D.~Madrigal Mart\'inez\\
        Instituto de F\'isica Te\'orica UAM/CSIC, Nicol\'as Cabrera 15, UAM, E-28049 Madrid, Spain\\
        E-mail: \email{josedaniel.madrigal@uam.es}}

\author{A. Sabio Vera\\
        Instituto de F\'isica Te\'orica UAM/CSIC, Nicol\'as Cabrera 15, UAM, E-28049 Madrid, Spain\\
        E-mail: \email{agustin.sabio@uam.es}}

\abstract{We discuss computational details
of our recent result, namely, the first derivation of the two-loop gluon Regge
  trajectory within the framework of Lipatov's high energy effective action.
  In particular,  we elaborate on the 
  direct evaluation of Feynman two-loop diagrams by using the Mellin-Barnes
  representations technique.
  Our result is in precise agreement with previous computations in the
  literature, providing this way a highly non-trivial test of the
  effective action and the proposed subtraction and renormalization
  scheme combined with our approach for the
  treatment of the loop diagrams.}

\FullConference{XXI International Workshop on Deep-Inelastic Scattering and Related Subject -DIS2013,\\
		22-26 April 2013\\
		Marseilles,France}

\begin{document}

\section{Introduction}

A particularly active field in perturbative QCD which addresses both fundamental
formal issues and phenomenological applications is the resummation program
of the high center-of-mass energy logarithms. The formalism was initiated almost forty
years ago by the Balitsky-Fadin-Kuraev-Lipatov (BFKL) equation \cite{BFKL1,BFKLNLO},
which emerged within the framework of high energy factorization.
Examples of
current applications of the high energy factorization to QCD phenomenology
can be found in the analysis of  dijets widely
separated in rapidity \cite{jets,Dusling:2012cg,Colferai:2010wu, Caporale:2012ih}, 
the studies of the transverse momentum dependent
parton distribution functions in the low $x$ region \cite{upd, Ellis:2008yp, Hentschinski:2012kr} and the
the study of observables in heavy ion collisions \cite{hic}. 
Their unifying factor is the factorization of
QCD scattering amplitudes in the limit of infinitely high center of
mass energies, combined with a resummation of large perturbative
logarithmic enhanced corrections.

A very powerful tool for the study of high energy scattering amplitudes is proving to 
be Lipatov's high
energy gauge invariant effective action \cite{LevSeff}, which attempts
to reformulate the high energy limit of QCD as an effective field
theory of reggeized gluons.  While the determination of the high
energy limit of tree-level amplitudes is well understood for a long
time within this framework \cite{Antonov:2004hh}, it was only
recently that progress in the calculation of loop corrections has been
achieved.  Starting in Ref.~\cite{quarkjet} and extended in
Refs.~\cite{gluonjet,Hentschinski:2011xg}, a scheme has been developed,
combining a subtraction procedure to avoid double counting and the
regularization and renormalization of high
energy divergences. This scheme then allowed to successfully derive
jet vertices for both quark and gluon initiated jets at NLO accuracy
from Lipatov's high energy effective action.
Within this formalism, we were able to address the more
challenging project we discuss here: the determination of the
two-loop gluon Regge trajectory~\cite{Chachamis:2012gh, review, Chachamis:2013hma}.
The gluon Regge trajectory provides an essential
ingredient in the formulation of high energy factorization and
reggeization of QCD amplitudes at NLO and was originally
derived in Refs.~\cite{Fadin:1996tb, Fadin:1995km} (see also 
relevant work in Refs.~\cite{Blumlein:1998ib, Korchemskaya:1996je,DelDuca:2001gu}).

Within Lipatov's effective action approach, to compute 
the gluon Regge trajectory at NLO, one has to calculate
two-loop corrections to the reggeized gluon propagator which, in other
words means computing two-loop self-energy Feynman diagrams. This
step was actually the main bottleneck of the project, for details, we refer the reader to the
original publications~\cite{Chachamis:2012gh, review, Chachamis:2013hma}.
Here, we will restrict ourselves to the discussion of the techniques
we used to evaluate the two-loop Feynman diagrams.

\section{The computation of the master integrals}

We focus on the gluonic contributions (no quark loops in the diagrams)
to the NLO gluon Regge trajectory which are technically more involved.
We remind the reader\footnote{For details, 
see Refs.~\cite{Chachamis:2012gh, review,Chachamis:2013hma}.} 
that we decompose momenta four-vectors 
not in terms of light-like vectors $n^{\pm}$ but in terms
of deformed light-like vectors, $n^-   \to n_a = e^{-\rho} n^+ + n^-$ 
and $n^+  \to n_b = n^+ + e^{-\rho} n^-$, where $\rho$ is an external
parameter used to regularize longitudinal divergencies 
and the amplitude has to be evaluated in the limit
$\rho \to \infty$. As it turns out, only a subset of diagrams can contribute, namely,
those that could be potentially enhanced by a factor $\rho^k$,  $k \geq 1$.

We have used the 
Mathematica package FIRE \cite{Smirnov:2008iw} which is an implementation of
the Laporta algorithm \cite{Laporta:2001dd}  to reduce the number and
complexity of the initial set of integrals into a smaller set of master integrals
through integration-by-parts identities
\cite{Chetyrkin:1981qh}. 
The generic two-loop master integral in our work can be represented\footnote{We use
dimensional regularization with $d = 4 + 2 \epsilon$.}
by 
\begin{align}
\label{eq:MASTER}
& \text{MI}[\,\alpha_1, \alpha_2,  \cdots,\alpha_9\,] = 
(\mu^4)^{-2 \epsilon}\iint\frac{d^dk}{(2\pi)^d}\frac{d^dl}{(2\pi)^d}\frac{1}{(-k^2-i0)^{\alpha_1}[-(k-q)^2-i0]^{\alpha_2}(-l^2-i0)^{\alpha_3} } 
\notag \\
& \times\frac{1}{
[-(l-q)^2-i0]^{\alpha_4}[-(k-l)^2-i0]^{\alpha_5}} \cdot \frac{1}{(-n_a\cdot k)^{\alpha_6}(-n_b\cdot k)^{\alpha_7}(-n_a\cdot l)^{\alpha_8}(-n_b\cdot l)^{\alpha_9}},
\end{align}
where $q$ is the momentum of the reggeized gluon whereas
$\xi=n_a^2=n_b^2=4e^{-\rho}$, $\delta=n_a\cdot n_b\sim 2$.

Dropping all terms that cannot give contributions to the NLO gluon Regge trajectory, 
we finally have to compute six master integrals, $\mathcal{A}$ to  $\mathcal{F}$, 
 which can be seen in Tab.~\ref{tab:masters},
each uniquely defined by the 
powers of its propagators,
along with a certain pre-coefficient associated to each one of them.

\renewcommand{\arraystretch}{2}
\begin{table}[th]
  \centering
  \begin{align*}
  \begin{array}[h]{rcl|c}
&&\text{master integral} & \text{coefficent} \\
\hline 
   {\cal A}   &\equiv &  \big[1,1,1,1,0,1,0,0,1\big]  & \displaystyle  c_{\mathcal{A}} = -\frac{\bm{q}^2}{2}\\ 
 {\cal B}   &\equiv & 
 \big[1,0,0,1,1,0,0,1,1\big] &  \displaystyle  c_{\mathcal{B}} = \frac{66+42\epsilon}{3+2\epsilon}\\ 
 {\cal C}   &\equiv & 
\big[1,1,1,1,1,1,0,0,1\big]  &  \displaystyle  c_{\mathcal{C}} = -(\bm{q}^2)^2\\ \displaystyle
 {\cal D}   &\equiv & 
[1,0,0,1,1,1,1,1,1\big]  & c_{\mathcal{D}} = -\bm{q}^2 \\
  {\cal E}   &\equiv &
\big[1,1,0,1,1,1,1,-1,1\big] & c_{\mathcal{E}} = -2{
\xi}\bm{q}^2\\
 {\cal F}   &\equiv & 
\big[1,1,1,1,1,1,-2,0,1\big] &     \displaystyle c_{\mathcal{F} = }-{
\xi}\bm{q}^2\\
  \end{array}
\end{align*}
  \caption{Master integrals and  pre-coefficients. 
  There is also a common overall factor  $(-2i\bm{q}^2)g^4N_c^2$.}
  \label{tab:masters}
\end{table}

To evaluate the master integrals we have used the Mellin-Barnes representations
technique (for a review see e.g. \cite{Smirnov:2006ry}).
For that purpose, we re-expressed the diagrams in terms
of Schwinger parameters so that we could apply
the basic Mellin-Barnes formula:
\begin{equation}\label{MB}
\begin{aligned}
\frac{1}{(X_1+\cdots+X_n)^\lambda}&=\frac{1}{\Gamma(\lambda)}\frac{1}{(2\pi i)^{n-1}}\int\cdots\int_{-i\infty}^{+i\infty} dz_2\cdots dz_n\prod_{i=2}^n X_i^{z_i}X_1^{-\lambda-z_2-\cdots-z_n}\\&\times\Gamma(\lambda+z_2+\cdots +z_n)\prod_{i=2}^n \Gamma (-z_i),
\end{aligned}
\end{equation}
where the integration contours are such that poles with a $\Gamma(\cdots+z_i)$ dependence are to the left of the $z_i$ contour and poles with a $\Gamma(\cdots-z_i)$ dependencies lie to the right of the $z_i$ contour.
The Mellin-Barnes representation of the generic master integral reads:
\begin{equation}\label{S}
\begin{aligned}
\mathcal{S}& {=\iint\frac{d^dk}{(2\pi)^d}\frac{d^dl}{(2\pi)^d}\frac{1}{[-k^2-i0]^{\alpha_1}[-(k-q)^2-i0]^{\alpha_2}[-l^2-i0]^{\alpha_3}[-(l-q)^2-i0]^{\alpha_4}}}\\
&  \times {\frac{1}{ [-(k-l)^2-i0]^{\alpha_5} (-n_a\cdot k-i0)^{\alpha_6}(-n_b\cdot k-i0)^{\alpha_7}( n_a\cdot l-i0)^{\alpha_8}(- n_b\cdot l-i0)^{\alpha_9}}}
\\
&=\frac{- \left(\bm{q}^2\right)^{d-{\alpha_{12345}}-\frac{\alpha_{6789}}{2}}}{2(4\pi)^d}
\prod_{i=1}^6 \int\frac{dz_i}{2\pi i} \Gamma(-z_i) \,  \frac{\Gamma\left(z_{1234}+{\alpha_{345}}+\tfrac{\alpha_{6789}}{2}-\frac{d}{2}\right)  \Gamma(-z_4+{\alpha_2}) }{ \prod_{j=1}^9\Gamma(\alpha_j) \Gamma(-2z_1)\Gamma(-2z_6) }
\\
&
\frac{  \Gamma\left(-z_{12345}-{\alpha_{345}}+\tfrac{\alpha_{6789}}{2}+\tfrac{d}{2}\right)\Gamma\left(-z_1+z_{2345}+{\alpha_{345}}-\tfrac{\alpha_{6789}}{2}-\tfrac{d}{2}\right)  \Gamma(-z_{3}+{\alpha_1})  }{\Gamma(-2z_2-z_{34}-{\alpha_{126789}}-2{\alpha_{345}}+2d)  }
\\
&
 \frac{\Gamma\left(-z_{23}-{\alpha_{2345}}-\tfrac{\alpha_{6789}}{2}+d\right)\Gamma\left(-z_{24}-{\alpha_{1345}}-\tfrac{\alpha_{6789}}{2}+d\right)\Gamma(z_{2345}-z_6+{\alpha_{3458}}-\tfrac{d}{2})}{\Gamma(2z_{234}+z_5+2{\alpha_{345}}+\alpha_{789}-d)}
\\
&
\frac{\Gamma(2z_{234}+z_5+2{\alpha_{345}}+\alpha_{89}-d)\Gamma(-z_{234}+z_6-{\alpha_{345}}+\tfrac{d}{2}) \Gamma\left(z_2+{\alpha_{12345}}+\tfrac{\alpha_{6789}}{2}-d\right)}{\Gamma(-z_5+\alpha_6)}
\\
& 
\frac{ \Gamma(-z_{23456}\! -\!{\alpha_{3458}}+\tfrac{d}{2})\Gamma(-2z_2-z_{34}-2\alpha_{34}  -{\alpha_{589}} +d) 
 \Gamma(z_{23}+{\alpha_3})   \Gamma(z_{24}+{\alpha_4})   }
{ 
 \Gamma(-{\alpha_{34589}}  +d)}     e^{-z_{16}\rho}    
\end{aligned}
\end{equation}
where $z_{ijk\ldots} = z_i + z_j + z_k + \ldots$ and $\alpha_{ijk\ldots} = \alpha_i + \alpha_j + \alpha_k + \ldots$. 

Once we substitute the actual values of the powers of the denominators 
for each master integral in Eq.~\ref{S}, we determine the integration contours
using the Mathematica package \texttt{MB.m} \cite{Czakon:2005rk} which
also serves as our working platform.
We then perform an asymptotic expansion
in $e^{-\rho}$ to obtain the leading behavior, using 
\texttt{MBasymptotics.m}~\cite{Cza}.
As a next step we resolve the structure of singularities
in $\epsilon$, again using  \texttt{MB.m} ( the package \texttt{MBresolve} \cite{Smirnov:2009up}
was also used) and finally we proceed to the evaluation of those terms
that are $\rho$-enhanced, making use at this stage also of the
routine \texttt{barnesroutines.m}~\cite{Kos}.

\section{Conclusions}
We have discussed the approach we used in~\cite{Chachamis:2013hma} to compute the two-loop 
Feynman integrals emerging in our calculation of the NLO gluon
Regge trajectory. We have used the Mellin-Barnes representations
technique which was proved adequate and which
we believe will be further used in future 
calculations within the framework of Lipatov's high energy effective action.

\section*{Acknowledgments}
We thank J. Bartels, V. Fadin and L. Lipatov for their constant support
with useful discussions.
We acknowledge support by the
Research Executive Agency (REA) of the European Union under 
the Grant Agreement number PITN-GA-2010-264564 (LHCPhenoNet),
the Comunidad de Madrid
through Proyecto HEPHACOS ESP-1473, and MICINN (FPA2010-17747),
the Spanish Government and EU ERDF funds 
(grants FPA2007-60323, FPA2011-23778 and CSD2007-00042 
Consolider Project CPAN) and by GV (PROMETEUII/2013/007). 
G.C. acknowledges support from Marie Curie actions (PIEF-GA-2011-298582).
M.H. acknowledges support from the U.S. Department of Energy under contract
number DE-AC02-98CH10886 and a ``BNL Laboratory Directed Research and
Development'' grant (LDRD 12-034).

\end{document}